\documentstyle[12pt,epsfig]{article}
\newcommand{\gev}{\; \hbox{GeV}}
\newcommand{\Gc}{\gamma_{ 5}}

\newcommand{\spazio}{\vphantom{$\Big( \Big)$}}

\newcommand{\deu}{(\delta^u_{23})}

\newcommand{\bxs}{ B \rightarrow X_s \ell^+\ell^-}
\newcommand{\gcenu}{ B \rightarrow X_c e \nu}
\newcommand{\ds}{\displaystyle}

\renewcommand{\a}{\alpha}
\renewcommand{\b}{\beta}

\renewcommand{\d}{\delta}

\newcommand{\g}{\gamma}
\newcommand{\la}{\lambda}
\newcommand{\G}{\Gamma}

\newcommand{\sw}{\sin^2{\theta_W}}
\newcommand{\as}{\alpha_{s}}
\newcommand{\bea}{\begin{eqnarray}}
\newcommand{\eea}{\end{eqnarray}}
\newcommand{\beq}{\begin{equation}}
\newcommand{\eeq}{\end{equation}}
\newcommand{\nn}{\nonumber}
\newcommand{\fr}{\frac}
\newcommand{\hl}{\hline}
\begin{document}
\begin{titlepage}
\vskip 1cm
\centerline{\Large $\mathbf B \rightarrow X_s \ell^+ \ell^-$  \bf decays in supersymmetry}
\vskip 1cm
\centerline{E. Lunghi, A. Masiero,}
\vskip 0.5cm
\centerline{ SISSA-ISAS, Via Beirut 2-4, Trieste, Italy and}
\centerline{INFN,  Sezione di Trieste, Trieste, Italy }
\vskip 0.5cm
\centerline{I. Scimemi}
\vskip 0.5cm
\centerline{Dep. de Fisica Teorica, Univ. de Valencia,}
\centerline{c. Dr.Moliner 50, E-46100, Burjassot, Valencia, Spain}
\vskip 0.5cm
\centerline{and}
\vskip 0.5cm
\centerline{L. Silvestrini}
\vskip 0.5cm
\centerline{Physik Department, Technische Universit{\"a}t M{\"u}nchen,}
\centerline{D-85748 Garching, Germany}

\bigskip
\rightline{Ref.~SISSA-17/99/EP}
\rightline{Ref.~TUM-HEP-346/99}
\rightline{Ref.~FTUV/99-30}
\rightline{IFIC/99-32}
\begin{abstract}
We study the semileptonic decays $B\rightarrow X_s e^+ e^-$, 
$B\rightarrow X_s \mu^+ \mu^-$ 
in generic supersymmetric extensions of the Standard Model.
SUSY effects are parameterized using the mass insertion approximation formalism
 and differences with the Constrained MSSM results are pointed out. Constraints on SUSY
contributions coming from other processes ({\it e.g.} $b\rightarrow s \g$) are taken into
account. Chargino and gluino contributions to photon and Z-mediated decays are
computed and non-perturbative  corrections are considered.
We find that the integrated branching ratios  and the asymmetries can
be strongly modified.
Moreover, the behavior of the differential Forward-Backward asymmetry
remarkably changes with respect to the  Standard Model expectation. 

\end{abstract}

\end{titlepage}
\newpage

\section{Introduction}
\label{sec:intro}
One of the features of 
 a general low energy supersymmetric (SUSY) extension of  
 the Standard Model (SM) is
the presence of a huge number of new parameters. FCNC and CP violating
 phenomena constrain strongly a big part of the new parameter space. 
However there is still room for significant departures from the SM
 expectations in this interesting class of physical processes.
It is worthwhile to check all these possibilities on the available data and 
on those processes that are going to be studied in the next future.
In this way it is possible to indicate where new physics effects can be 
 revealed as well as to establish criteria for model building.

In this work we want to investigate the  relevance
 of new physics effects in the 
semileptonic inclusive decay $B \rightarrow X_s
\ell^+ \ell^-$.
This decay is quite suppressed in the Standard Model;
however, new $B$-factories should  reach the precision requested by the
 SM prediction~\cite{babar} and an estimate of all possible new contributions to
 this process is compelling.

Semileptonic charmless $B$ decays have been deeply studied.
The dominant perturbative SM contribution has been evaluated  in 
ref.~\cite{grin} and later two loop QCD corrections have been 
provided~\cite{bura,bura1}. The contribution due to  $c\bar c$ resonances 
 to these results are included
in the papers listed in ref.~\cite{tram}.
Long distance corrections   can  have a different  origin  according to
the value of the dilepton invariant mass one considers.
$O(1/m_b^2)$ corrections have been first calculated  
in ref.~\cite{falk} and recently  corrected in ref.~\cite{alih,buch}. 
Near the peaks, non-perturbative contributions generated by $c\bar c$ resonances  by means of
resonance-exchange models have been  provided in ref.~\cite{alih, krug}.
Far from the resonance region, instead, ref.~\cite{rey} (see also ref.~\cite{rupa}) 
estimate $c\bar c$ long-distance effects
using a heavy quark expansion in inverse powers of the
 charm-quark mass ($O(1/m_c^2)$ corrections).

An analysis  of the SUSY contributions
 has been presented in refs.~\cite{bert}--\cite{goto}  
 where the authors estimate
 the contribution of the Minimal Supersymmetric
Standard Model (MSSM). 
They consider first  a universal soft supersymmetry breaking sector
at the Grand Unification scale (Constrained MSSM) 
and then partly relax this universality condition.
In the latter case they  find that there can be a substantial difference between
the SM and the SUSY results in the Branching Ratios and in the
 forward--backward asymmetries. One of the reasons of this enhancement
is that the Wilson coefficient $C_7(M_W) $ (see section~\ref{sec:opba} for a
precise definition) can change sign 
with respect to the SM in some region of the parameter space
 while respecting constraints coming from $b\rightarrow s \g$.
The recent measurements of   $b\rightarrow s \g$~\cite{cleo}
 have narrowed the window
of the possible values of  $C_7(M_W)$ and in particular a sign change of this
coefficient is no more allowed in the Constrained MSSM framework.
Hence, on one hand it is worthwhile considering $B \rightarrow X_s \ell^+
 \ell^-$ in a more general SUSY framework then just the Constrained MSSM, and, on the
 other hand, the above mentioned new results prompt us to a
 reconsideration of the process.
 In reference~\cite{korea} the possibility
 of new-physics effects coming from gluino-mediated FCNC is studied.
Effects of SUSY phases in models with heavy first and second
 generation sfermions have been recently discussed in ref.~\cite{ko}.

We consider all possible contributions to charmless semileptonic $B$ decays 
 coming from chargino-quark-squark and gluino-quark-squark interactions
and we analyze both Z-boson and photon mediated decays.
Contributions coming from penguin and box diagrams are taken into account;
moreover, corrections to the mass insertion approximation (see below)
 results due to a light $\tilde t_R$ are considered.
A direct comparison between the SUSY and the SM contributions to
the Wilson coefficients  is performed.
Once   the constraints on mass insertions are established,
we find that in generic SUSY models
 there is still  enough room  in order to see large deviations from
 the SM expectations for branching ratios and asymmetries. 
For our final computation of physical observables
 we consider NLO QCD evolution of the coefficients and 
non-perturbative corrections ($O(1/{m_b^2}),\ O(1/{m_c^2}),$...),   
each in its proper range of the dilepton invariant mass.

Because of the presence of so many unknown parameters (in particular in the
scalar mass matrices) which enter in a quite complicated way in
the determination of the mass eigenstates and of the various mixing matrices
it is very useful to adopt the so-called 
``Mass Insertion Approximation''(MIA)~\cite{hall}. 
In this framework one chooses a basis for fermion and sfermion states
in which all the couplings of these particles to neutral gauginos are
 flavor diagonal. Flavor changes in the squark sector are provided
 by the non-diagonality of the  sfermion propagators.
The pattern of flavor change is then given by the ratios 
\beq
(\delta^{f}_{ij})_{AB}= 
\fr{(m^{\tilde f }_{ij})^2_{AB}}{M_{sq}^2} \ ,
\eeq
where $ (m^{\tilde f }_{ij})^2_{AB}$  are the off-diagonal elements of the
$\tilde f=\tilde u,\tilde d $ mass squared matrix that
 mixes flavor $i$, $j$ for both left- and
right-handed scalars ($A,B=$Left, Right) and  $M_{sq}$ is the
average squark mass (see {\it e.g.}~\cite{gabb}).
The sfermion propagators are expanded in terms of the $\delta$s
and the contribution of the first two terms of this expansion are considered.
The genuine SUSY contributions to the Wilson coefficients will be simply
proportional to the various $\delta$s and a keen analysis of 
the different Feynman diagrams involved will allow us to isolate the few
insertions really relevant for a given process. 
In this way we see that only a small number of the new parameters is
involved and a general SUSY analysis is made possible.
The hypothesis regarding the smallness of the $\delta$s and so the
reliability of the approximation can then be checked {\it a posteriori}.

Many of these $\delta$s are strongly constrained
by FCNC effects~\cite{gabb,hage,gab} or by
vacuum stability arguments~\cite{casa}.
Nevertheless  it may happen  that such limits are not strong enough to prevent
large contributions to some rare processes.
For instance it has been recently found in ref.~\cite{cola} that
the off-diagonal squark mass matrix elements can enhance rare kaon decays
by roughly an order of magnitude with respect to the SM result.

The paper is organized as follows.
In sect.~\ref{sec:opba} we define the operator basis, the basic formulae
for the BR, the Forward--Backward asymmetry 
and the non-perturbative corrections.
Sect.~\ref{sec:chains} and sect.~\ref{sec:gluins} treat chargino and
gluino    contributions in the mass insertion approximation.
The light  $\tilde t_R$ corrections are presented in sect.~\ref{sec:lightstop}.
Constraints on $\delta$s are discussed in sect.~\ref{sec:deltas} and final results
and conclusions are drawn in sect.~\ref{sec:results}
 and \ref{sec:conclusions}.

%___________________________________________________________________________
\begin{figure}
\begin{center}
\hspace*{-5cm}
\epsfig{file=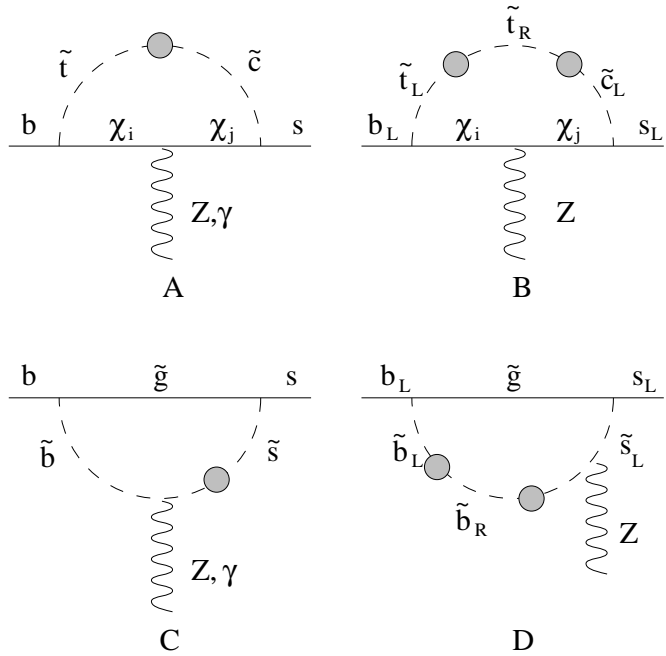,width=0.6\linewidth,angle=270}
\caption[]{\it Some of the relevant penguin diagrams for $b\rightarrow
s \ell^+ \ell^-$. The bubble
indicates a Mass Insertion. Diagrams  A,B are based on chargino interaction.
Diagrams C,D  consider gluino interactions.
 }
\protect\label{fig:sfigs}
\end{center}
\end{figure}
%___________________________________________________________________________
%___________________________________________________________________________
\begin{figure}
\begin{center}
\vspace*{1cm}
\epsfig{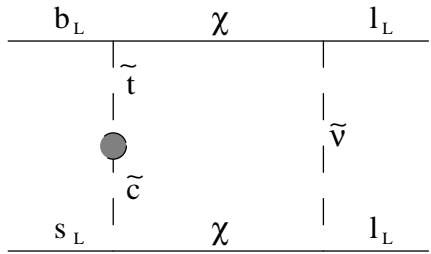}
\vspace*{-1cm}
\caption[]{\it Relevant box diagram for $b\rightarrow s \ell^+
\ell^-$. The bubble
indicates a Mass Insertion. }
\protect\label{fig:box}
\end{center}
\end{figure}
%___________________________________________________________________________
\newpage
\section{Operator basis and general framework}
\label{sec:opba}
The effective Hamiltonian for the decay $ B \rightarrow X_s \ell^+\ell^-$
in the SM and in the MSSM is given by (neglecting the
small contribution proportional to $K^*_{us} K_{ub}$) 

\begin{equation}
{\cal H}_{\rm eff} = - {4 G_F \over \sqrt{2}} K^*_{ts} K_{tb} \left[
                     \sum_{i=1}^8 C_i (\mu) Q_i + 
                     {\a \over 4 \pi} \sum_{i=9}^{10} \tilde{C}_i (\mu) Q_i
                     \right],
\end{equation}
where 
\begin{eqnarray}
Q_1 &=& \bar{s}_{L\a} \g_\mu b_{L\a} \bar{c}_{L\b} \g^\mu c_{L\b}, \nn \\ 
Q_2 &=& \bar{s}_{L\a} \g_\mu b_{L\b} \bar{c}_{L\b} \g^\mu c_{L\a}, \nn \\ 
Q_3 &=& \bar{s}_{L\a} \g_\mu b_{L\a} \sum_{q=u,..,b} \bar{q}_{L\b} \g^\mu q_{L\b}, \nn \\ 
Q_4 &=& \bar{s}_{L\a} \g_\mu b_{L\b} \sum_{q=u,..,b} \bar{q}_{L\b} \g^\mu q_{L\a}, \nn \\ 
Q_5 &=& \bar{s}_{L\a} \g_\mu b_{L\a} \sum_{q=u,..,b} \bar{q}_{R\b} \g^\mu q_{R\b}, \nn \\ 
Q_6 &=& \bar{s}_{L\a} \g_\mu b_{L\b} \sum_{q=u,..,b} \bar{q}_{R\b} \g^\mu q_{R\a}, \nn \\ 
Q_7 &=&\fr{e}{16 \pi^2}m_b\bar{s}_L\sigma^{\mu\nu}b_R F_{\mu\nu}, \nn\\
Q_8 &=&\fr{g_s}{16 \pi^2}m_b \bar{s}_L T^a\sigma^{\mu\nu}b_R G_{\mu\nu}^a, \nn\\
Q_9 &=&(\bar{s}_L\g_\mu b_L)\bar{l}\g^\mu l, \nn \\
Q_{10} &=&(\bar{s}_L\g_\mu b_L)\bar{l}\g^\mu \g_5 l\ , 
\end{eqnarray}
$K$ is the CKM-matrix and $\ds q_{L(R)}= {(1\mp \Gc) \over  2} \;q$.
This Hamiltonian is known at next-to-leading order both in the SM~\cite{bura,bura1} and
in the MSSM~\cite{cho,goto}.
We  find that the most general low-energy SUSY Hamiltonian 
also contains the operators
 \bea
Q_7^\prime&=&\fr{e}{8 \pi^2}m_b\bar{s}_R\sigma^{\mu\nu}b_L F_{\mu\nu},\nn\\
Q_9^\prime&=&(\bar{s}_R\g_\mu b_R)\bar{l}\g^\mu l, \nn \\
Q_{10}^\prime&=&(\bar{s}_R\g_\mu b_R)\bar{l}\g^\mu \g_5 l .
\label{eq:ope}
\eea
However  it is shown in following sections that the contribution
of these operators is negligible and so they are not
considered in the final discussion of physical quantities.
SUSY contributions to other operators are negligible because they 
influence our observables  at an higher perturbative order. 

With these definitions the differential branching ratio and 
the forward-backward asymmetry  can be written as
\begin{eqnarray}
R(s) &\equiv& {1\over \G (\gcenu)} {{\rm d} \ \G (\bxs) \over {\rm d}  s} \nn \\
&=& {\a^2 \over 4 \pi^2} \left| K_{ts}^* \over K_{cb} \right| {(1-s)^2 \over f(z) k(z)} \times \nn \\
& & \left[ (1+2s)\left( |\tilde{C}_9^{\rm eff}(s)|^2 + |\tilde{C}_{10}|^2 \right)
 +4(1+{2\over s}) |C_7|^2 + 12 {\rm Re} \left[ C_7^* \tilde{C}_9^{\rm eff}(s) \right] \right],     
\label{eq:br0} \\
A_{FB}(s) & \equiv & {\ds \int_{-1}^1 {\rm d} \cos{\theta} \; {\ds{\rm d}^2 
		     \G (\bxs)\over
                     \ds {\rm d} \cos{\theta} \; {\rm d} s} \; {\rm Sgn} (\cos{\theta})    
                     \over
                     \ds \int_{-1}^1 {\rm d}\cos{\theta} \; {\ds {\rm d}^2 \G (\bxs)\over
                     \ds {\rm d}\cos{\theta} \; {\rm d} s}}  \nn \\
          & = & - \fr{3 {\rm Re} \left[ \tilde{C}_{10}^* \left(
                    s \ \tilde{C}_9^{\rm eff}(s)  + 2 C_7 \right)
		     \right]}{ \ds (1+2s)\left( |\tilde{C}_9^{\rm
		     eff}(s)|^2 + |\tilde{C}_{10}|^2 \right)
                 +4(1+{2\over s}) |C_7|^2 + 12 {\rm Re} \left[ C_7^*
		     \tilde{C}_9^{\rm eff}(s) \right] } 
\label{eq:afb0} 
\end{eqnarray}
where $s=(p_{\ell^+}+p_{\ell^-})^2 /m_b^2$, $\theta$ is
the angle between the positively charged lepton and the B flight
direction in the rest frame of the dilepton system, $f(z)$ and $k(z)$ 
are the phase space factor and 
the QCD correction factor ($z=m_c / m_b$) that enter $\G (\gcenu)$ and
can be found in refs.~\cite{bura1,kim}. 
$\tilde{C}_9^{\rm eff}(s)$  includes all the contributions of the operators
$Q_1-Q_6$ and $Q_8$  and its complete definition
for  the SM and MSSM can be found again in refs.~\cite{bura,bura1,cho}.

In the literature the energy asymmetry is also considered~\cite{cho}
but it is easy to show that these two kind of asymmetries are
completely equivalent; in fact a configuration in the dilepton
c.m.s. in which $\ell^+$ is scattered in the forward direction
kinematically implies $E_{\ell^+} < E_{\ell^-}$ in the B rest frame
(see for instance ref.~\cite{alih}).

It is worth underlying that integrating the differential asymmetry
given in eq.~(\ref{eq:afb0}) we do not obtain the global
Foward--Backward asymmetry which is by definition:
\begin{equation}
{N(\ell^+_\rightarrow) - N(\ell^+_\leftarrow)
           \over N(\ell^+_\rightarrow) + N(\ell^+_\leftarrow)} =
 {\ds \int_{-1}^1 {\rm d} \cos{\theta} \int {\rm d} s\; {\ds{\rm d}^2 
		  \G (\bxs)\over
                     \ds {\rm d} \cos{\theta} \; {\rm d} s} \; {\rm Sgn} (\cos{\theta})    
                     \over
           \ds \int_{-1}^1 {\rm d}\cos{\theta} \int{\rm d} s \; {\ds {\rm d}^2 \G (\bxs)\over
                     \ds {\rm d}\cos{\theta} \; {\rm d} s}} 
\label{intafb2}
\end{equation}
where $\ell^+_\rightarrow$ and $\ell^+_\leftarrow$ stand respectively for
leptons scattered in the forward and  backward direction.

\noindent To this extent it is useful to introduce the following quantity
$$
\overline A_{FB} (s)  \equiv  {\ds \int_{-1}^1 {\rm d} \cos{\theta} \; {\ds{\rm d}^2 
		     \G (\bxs)\over
                     \ds {\rm d} \cos{\theta} \; {\rm d} s} \; {\rm Sgn} (\cos{\theta})    
                     \over
           \ds \int_{-1}^1 {\rm d}\cos{\theta} \int {\rm d} s \; {\ds {\rm d}^2 \G (\bxs)\over
                     \ds {\rm d}\cos{\theta} \; {\rm d} s}}   
$$
\begin{equation}
 =  \fr{-3 {\rm Re} \left[ \tilde{C}_{10}^* \left(
                    s \ \tilde{C}_9^{\rm eff}(s)  + 2 C_7 \right)
		     \right] (1-s)^2}{                      
                 \ds \int {\rm d}s (1-s)^2 \left[ (1+2s)\left(
		     |\tilde{C}_9^{\rm eff}(s)|^2 + |\tilde{C}_{10}|^2 \right)
                 +4(1+{2\over s}) |C_7|^2 + 12 {\rm Re}\left[ C_7^*
		     \tilde{C}_9^{\rm eff}(s)\right] \right] } 
\label{eq:afb20} 
\end{equation}	
 whose integrated value is given by eq.~(\ref{intafb2}).

Eqns.~(\ref{eq:br0}) and (\ref{eq:afb0}) have been corrected in order
 to include several non-perturbative effects.
First of all $O(1/m_b^2)$ effects have been estimated by \cite{alih},
\bea
\d_{1/m_b^2} R(s)&=&\fr{3 \la_2}{2 m_b^2}
\left[\fr{\a^2}{4 \pi^2}\left|\fr{K_{ts}^*}{K_{cb}}\right|^2
{1\over f(z) k(z)}\Big( (1-15 s^2+10 s^3) 
\left( |\tilde{C}_9^{\rm eff}(s)|^2 + |\tilde{C}_{10}|^2 \right)-
 \Big.\right.
\nn \\
& & \left.\Big.
4(6+3 s -5 s^3) {|C_7|^2\over s}-(5 +6 s -7 s^2)
 {\rm Re} \left[ 
C_7^* \tilde{C}_9^{\rm eff}(s)\right] \Big) 
+{g(z)\over f(z) } R(s) \right],
\label{eq:brmb} \\
\d_{1/m_b^2} A_{FB}(s)&=&\fr{3 \la_2}{2 m_b^2} 
{(1-s)^{-2} {\rm Re}\left[ 
\tilde{C}_{10}^* \left[
 s (9 + 14 s -15 s^2) \tilde{C}_9^{\rm eff}(s)+
2 (7+10 s-9 s^2) C_7\right]\right]
\over
 (1+2s)\left( |\tilde{C}_9^{\rm eff}(s)|^2 + |\tilde{C}_{10}|^2 \right)
    +4(1+{2/ s})
 |C_7|^2 + 12 {\rm Re} \left[ C_7^* \tilde{C}_9^{\rm eff}(s) \right] }+
 \nn \\
& &
A_{FB}(s)\left( \fr{3 \la_2}{2 m_b^2} {g(z)\over f(z)} 
+\fr{4 \la_1}{3 m_b^2}\fr{s}{(1-s)^2}
-\fr{\d_{1/m_b^2} R(s)}{R(s)} \right),
\label{eq:afbmb}
\eea
 where
\bea
g(z)&=&
3-8 z^2 +24 z^4 -24 z^6 +5  z^8+24 z^4 \log{[z]}\ .
\nn
\eea
$\la_1$ and $\la_2$ are the  two parameters that appear in
the Heavy Quark Expansion (HQE).
While the value of $\la_2$  is quite well-established, 
($\la_2=(M_{B^*}^2-M_B^2)/4)$,  $\la_1$ is not yet well known.
In  ref.~\cite{ball} $\la_1$ is estimated  as 
$\la_1=-0.52 \pm 0.12 $ GeV$^2$ and in ref.~\cite{nela}
$\la_1=-0.10 \pm 0.05 $ GeV$^2$.
In what follows we consider the  weighted average of the two results
$\la_1=-0.16 $ GeV$^2$.
As was pointed out by the authors of 
ref.~\cite{alih} these  corrections are no more valid  near the endpoint region,
 $s\rightarrow 1$,
where they diverge because of the breaking down of the HQE. 
Following some recent analyses 
we have stopped the BRs corrections given
in eq.~(\ref{eq:brmb}) at $s=0.78$ (see ref.~\cite{buch}) 
and the ones in eq.~(\ref{eq:afbmb}) at $s=0.4$ (see ref.~\cite{alih}).

In order to   account for  the corrections to the parton model
approximation in the high $s$ region ref.~\cite{alih} and ref.~\cite{buch}
 adopt two different approaches.
The former  considers  a Fermi-motion model and the latter invokes
the Heavy Hadron Chiral Perturbation Theory (HHChPT).
A discussion about the usefulness 
of the   Fermi-motion model for semileptonic charmless
$B$-decays  is beyond the scope  of this paper.
In order to have a model independent description of the high energy
region of the spectrum we have considered the HHChPT corrections.

For $0.88<s<s_{max}=(M_B-M_K)^2/m_b^2=0.99$ 
the branching ratio is dominated by the exclusive decays
 $B\rightarrow K \ell^+ \ell^-$  and $B\rightarrow K \pi \ell^+ \ell^-$;
in ref.~\cite{buch} is shown that the contribution of the latter is
completely negligible. In the following we report the expression of the 
branching ratio, valid in the interval of the spectrum given above, 
computed in the HHChPT framework 
\bea
\hspace{-1cm}
R(s)_{\rm Ch}  &=& \fr{\tau(B_d)}{R(B\rightarrow X_c e\nu)}
\fr{G_F^2 M_B^5}{192 \pi^3} |K_{ts}^* K_{tb}|^2 \fr{\a^2}{4 \pi^2} 
\fr{f_1(s\ m_b^2/M_B^2 )}{2}
\times  \nn \\
& &  \left\{
f_+^2(s) \left( 
\left|\tilde{C}_9^{\rm EP} \right|^2+
\left|\tilde{C}_{10} \right|^2\right)+
a_T^2(s) m_b^2 \left|C_7\right|^2-
2 f_+(s) a_T(s) m_b {\rm Re} \left[C_7^* \tilde{C}_9^{\rm EP}\right]
\right\}
\label{eq:brch} 
\eea
where
\bea
a_T&=& \fr{g f_B}{f_\pi}\fr{1}{v\cdot p_K +(M_{B^*}-M_B)+(M_{B_s}-M_B)}
\eea 
and the definitions  of $f_1,\ f_+, \tilde{C}_9^{\rm EP}, \
v\cdot p_K$ can be
found in ref.~\cite{buch}. Moreover, we have put g=0.5 according to
the theoretical estimate given in ref.~\cite{cboni}.
In the intermediate region $0.78<s<0.88$ 
we have interpolated the  obtained results.
The form factors $f_+$ and $a_T$ can be computed also using other 
methods ( QCD Sum Rules~\cite{sum}, Ligh Cone QCD Sum
Rules~\cite{light}, QCD relativistic potential model~\cite{pote})
but the HHChPT approach is preferable in
the endpoint region of the spectrum. 

The asymmetries receive no contribution from the single kaon mode $B\rightarrow K \ell^+ \ell^-$
and the endpoint of their spectrum is fixed, instead,  by $B\rightarrow  K \pi \ell^+ \ell^-$ at
$s=(M_B-M_K-M_{\pi})^2/m_b^2=0.93$. In the region 
 $0.7<s<0.93$ we use the parton model
result because these asymmetries have not been computed yet in the 
HHChPT framework.

Finally, the following $O(1/m_c^2)$ corrections occur for $s<0.2$ (see
 {\it e.g.}~\cite{rey})
\bea
\d_{1/m_c^2}R(s)&=& -\fr{8\la_2}{9 m_c^2} C_2
 {\a\over 4 \pi^2} \left| K_{ts}^*\over K_{cb}\right|^2
\fr{(1-s)^2}{f(z) k(z)} \times \nn \\
& & 
{\rm Re}\left[ F(s) \left( C_7^* \fr{1+6 s-s^2}{s}+  \tilde{C}_9^{\rm eff* }(2+s)\right)
\right],
\label{eq:brmc} \\
\d_{1/m_c^2}A_{FB}(s)&=&\fr{3\la_2}{9 m_c^2} C_2 (1+ 3 s)\times \nn \\
& &
{{\rm Re}\left[F(s) \tilde{C}_{10}^*\right]\over
(1+2s)\left( |\tilde{C}_9^{\rm eff}(s)|^2 + |\tilde{C}_{10}|^2 \right)
 +4(1+{2/ s})
 |C_7|^2 + 12 {\rm Re} \left[ C_7^* \tilde{C}_9^{\rm eff} \right] }.
\label{eq:afbmc}
\eea

The $O(1/m_b^2)$ and $O(1/m_c^2)$ corrections to $\overline{A}_{FB}$ can be easily computed
because of the following relation
\begin{equation}
\overline{A}_{FB} = A_{FB} { R(s)\over \int {\rm d} s \ R(s)}.
\end{equation}

All the effects coming from  the mass insertion 
approximation  can be included in 
  formulae~(\ref{eq:br0}-\ref{eq:afbmc}) writing 
the coefficients $C_7$, $\tilde{C}_9^{\rm eff}(s)$, $\tilde{C}_{10}$ as 
\bea
C_7&=& C_7^{SM}+C_7^{Diag}+C_7^{MI}, \nn \\
\tilde{C}_9^{\rm eff}(s)&=& (\tilde{C}_9^{\rm eff}(s))^{SM}+
                          (\tilde{C}_9^{\rm eff})^{Diag}+
                          (\tilde{C}_9^{\rm eff})^{MI}, \nn \\
\tilde{C}_{10}&=&\tilde{C}_{10}^{SM}+\tilde{C}_{10}^{Diag}+\tilde{C}_{10}^{MI}
\label{eq:cstr}
\eea
where all the contributions are evaluated at the $M_B$ scale and 
the various $C_i^{Diag}$ summarize all the contributions coming from graphs
including SUSY Higgs bosons and sparticles in the limit in which we
neglect all the mass insertion contributions (they would be the only
SUSY diagrams if the scalar mass matrices were diagonalized by the same
rotations as those needed by the fermions). The explicit expressions
for $C_i^{Diag}$ can be found in ref.~\cite{cho}.

The Feynman diagrams with MI  relevant for 
$b \rightarrow s \ell^+ \ell^-$ are drawn in 
figs~\ref{fig:sfigs}-\ref{fig:box}.  We have considered  gluino-like and chargino-like
contributions  with both single and double mass insertions.

Both photons and Z-bosons can mediate  the decay.
Usually one finds that Z-boson contributions are dominant in those
graphs where an ``explicit $SU(2)_L$ breaking'' is provided, {\it i.e.},
both Left and Right squarks are present in the same loop.
In the latter  cases the photon  cannot feel any gauge-symmetry-breaking
and its contribution to the Wilson
coefficients is suppressed by a factor $m_Z^2/M_{sq}^2$ with respect
to the Z-boson one.
For $M_{sq}\sim 300$GeV, this factor amounts to about an order of magnitude.
On the other hand if the graph does not give any $SU(2)_L$ explicit
breaking we are in the opposite situation and
the Z--boson contribution is suppressed by a factor $m_b^2/m_Z^2\sim 3\cdot 10^{-3}$.
Moreover, a general feature
 of $\g$-mediated four fermion contributions is that,
for high average squark masses,
they decouple much faster  than in the Z-boson case.
This can be understood simply using dimensional arguments.
While Wilson coefficients for Z-boson mediated four-fermion interactions
are proportional to $\Delta /(m_Z^2 M_{sq}^2)$, the same coefficients must be proportional
to $\Delta/M_{sq}^4$ for the $\g$ ($\Delta$ here is a generic
off--diagonal element of the sparticle squared mass matrices and it
cannot rise as fast as $M_{sq}^2$ for high values of $M_{sq}$). Thus photon  graphs can  compete with
Z-boson graphs if the sparticle spectrum is not too heavy.

Finally the value of the
physical constants we use is reported in table~\ref{tab:cost}.
\begin{table}
\begin{center}
\begin{tabular}{||c|c||}
\hl 
$m_t$ & 173.8 GeV \\
$m_b$ & 4.8 GeV \\
$m_c$ & 1.4 GeV \\
$m_s$ & 125 MeV \\
$M_B$ & 5.27 GeV\\
$\a_s(m_Z)$ & .119 \\
$1/\a_{el}(m_Z)$ &128.9 \\
$\sin^2 \theta_W$ & .2334 \\  \hl 
\end{tabular}
\caption{Central values of physical constants used in the phenomenological analysis}
\label{tab:cost}
\end{center}
\end{table}

In the following subsections we describe in detail the 
 contributions
of each graph of figs.~\ref{fig:sfigs}-\ref{fig:box}.

\section{Chargino interactions}
\label{sec:chains}

In the weak eigenstates basis the chargino mass matrix
is given by
\begin{equation}
M_\chi =\left(
\begin{array}{cc}
M_2 & \sqrt{2}M_W \sin \beta \\
 \sqrt{2}M_W \cos \beta &\mu
\end{array}
\right)\ ,
\label{eq:char}
\end{equation}
where the index 1 of rows and columns refers to the the wino state,
the index 2 to the higgsino one, $\mu$ is the Higgs quadratic coupling
and $M_2$ the soft SUSY breaking wino mass. 
In order to define the mass eigenstates the unitary matrices
$U$ and $V$ which diagonalize $M_{\chi}$ are introduced,
\beq
{\rm diag}(M_{\chi_1},M_{\chi_2})= U^* M_{\chi} V^+  \ . \nn
\end{equation}
After the rotation to mass eigenstates it is always possible to speak
of wino-quark-squark or higgsino-quark-squark  interactions.
In order to identify the wino and higgsino states from the chargino ones it is 
sufficient to pick up the right elements from the $U$ and $V$ matrices.
To be clear we write them explicitly for the cases of interest
in the super-CKM basis. The wino-quark-squark, $d_L\tilde{W}\tilde{u}$,
   vertex is ($d$ and $\tilde u$ are a generic
 down-quark and up-squark)
\beq i g_2
 \sum_{j,k=1,2,3} K_{kj}^*
\sum_{a=1,2} V_{a1} \bar{d}_{jL} \chi _a {\tilde u}_{kL}
+{\rm h.c.} \ ,
\eeq
and the higgsino-quark-squark, $d \tilde{H}\tilde{u}$,  vertex is
\beq i
\sum_{j,k=1,2,3} K_{kj}^* \sum_{a=1,2} \left[ 
(\lambda^u)_k V_{a2} \; \bar d_{jL} \chi _a {\tilde u}_{kR} -
(\lambda^d)_k U_{a2}^* \; \bar d_{jR} \chi _a {\tilde u}_{kL}
\right] + {\rm h.c.} 
\label{eq:hqs}
\eeq
where $K$ is the CKM-matrix, $g_2= e/ \sin \theta_W$, 
$\lambda^u = {\ds \sqrt{2} M^u \over
\ds v \sin{\b}} = {\ds M^u g_2 \over \ds \sqrt{2} \sin{\b} M_W}$  and 
$\lambda^d = {\ds \sqrt{2} M^d \over \ds v \cos{\b}}$ 
are the Yukawa matrices for the up-- and down--quarks. 

Chargino graphs can contribute to the decay via both single and double
insertions (see figs.~\ref{fig:sfigs}-\ref{fig:box}).
The  double insertion is particularly convenient if the corresponding
$\d$s are not very constrained~\cite{cola}.
In the following subsections we examine both  the cases.
In the case of a single insertion approximation both $\g$- and Z- mediated
decays are considered. 

In all what follows our results for the integrals are written
in terms of the functions
\beq
P_{ijk}(a,b)\equiv 
\int_0^1 d x \int_0^1 d y \fr{y^i (1-y)^j}{(1 - y + a x y  + b (1-x) y
 )^k}\ .
\eeq
To get a feeling with numbers
it is sufficient to say that for $a=b=1$,
\beq
 P_{ijk}(1,1)=\beta_E[1+i,1+j]
\label{eq:euler}
\eeq
where  $\beta_E$ is the Euler  $\beta$-function.

\subsection{Single mass insertion -- $Z$}

The Z-boson mediated decay can proceed  in two ways  depending on the
type of chargino-quark-squark vertices we consider.
If an explicit $SU(2)_L$ breaking  on the
squark line of fig.~\ref{fig:sfigs}A is required we must take 
 both an higgsino  and a wino vertex.
In this way we  get a contribution to the Wilson coefficients
\bea
&&- \fr{C_9}{1-4\sw}=C_{10}=  
(\d_{23}^u)_{LR} {\lambda_t \over g_2} {K_{cs}^*\over K_{ts}^* } 
{1\over 4 \sw}
\sum_{i,j=1,2} V_{i1} V_{j2}^* \times \nn \\
 &&
\left\{U^*_{i1} U_{j1} \sqrt{x_{i} x_{j}} P_{112}(x_i,x_j) 
 + V^*_{i1} V_{j1} P_{111}(x_i,x_j)-
\fr{1}{2} \d_{ij} P_{021}(x_i,x_j)
\right\} \ ,
\label{eq:higw}
\eea
 where $x_i=M_{\chi_i}^2/M_{sq}^2$.
This diagram, however, is exactly null in the limit in which 
$U$, $V$ approximate the identity matrix and so it is  
negligible for high $M_2$.

With two wino-quark-squark vertices we obtain
\bea
&&-\fr{C_9}{1-4\sw}=C_{10}= 
-(\d_{23}^u)_{LL} {K_{cs}^*\over K_{ts}^*}{1\over 4 \sin^2{\theta_W}}
 \sum_{i,j=1,2} V_{i1} V_{j1}^* \times \nn \\
 &&
\left\{
U^*_{i1} U_{j1}
\sqrt{x_{i} x_{j}} P_{112}(x_i,x_j) 
 + V^*_{i1} V_{j1} P_{111}(x_i,x_j)-
 \d_{ij} P_{021}(x_i,x_j)
\right\}
\label{eq:winw}
\eea
where we have retained only the contribution which arises because of
the explicit $SU(2)_L$ breaking (with a double wino--higgsino mixing
in the wino line); in fact eq.~(\ref{eq:winw}) is null in the limit 
of diagonal chargino mass matrix.

Graphs with two higgsino-quark-squark vertices are suppressed  with
respect to these ones by  Yukawa or CKM factors.

\subsection{Single mass insertion -- $\gamma$}

The contributions of the $\g$ penguin with two wino vertices are
\bea
C_7&=&-(\d^u_{23})_{LL} \fr{M^2_W}{M_{sq}^2 }\fr{1}{3} 
   {K_{cs}^* \over K_{ts}^*} 
   \sum_{i=1,2} V_{i1} V^*_{i1} \left\{\fr{3}{2} P_{222}(x_i,x_i)+
    P_{132}(x_i,x_i)\right\}\ , 
\nn \\
C_9&=&-(\d^u_{23})_{LL}  \fr{M^2_W}{M_{sq}^2 }\fr{2}{3}
{K_{cs}^*\over K_{ts}^* }\times\nn \\
   & &\sum_{i=1,2} V_{i1} V^*_{i1} \left\{ P_{312}(x_i,x_i)+
   \fr{1}{3} P_{042}(x_i,x_i)+ x_i P_{313}(x_i,x_i) \right\}, 
\nn \\
C'_7&=&-(\d^u_{23})_{LL} \fr{M^2_W}{M_{sq}^2 }\fr{1}{3}
     {K_{cs}^* \over K_{ts}^*} \fr{m_s}{m_b}
    \sum_{i=1,2} V_{i1} V^*_{i1} \left\{\fr{3}{2} P_{222}(x_i,x_i)+
    P_{132}(x_i,x_i)\right\}\ . 
\eea

The contributions of the $\g$ penguin with an higgsino and a wino vertices are
\bea
C_7&=& \fr{M^2_W}{M_{sq}^2 }   
   {K_{cs}^* \over K_{ts}^*} 
   \sum_{i=1,2} \left[
    V_{i2}^* V_{i1}  {\lambda_t \over g_2} 
    \left\{\fr{1}{2} P_{222}(x_i,x_i)+{1\over 3}
    P_{132}(x_i,x_i)\right\} (\d^u_{23})_{LR} + \right. \nn \\
&&  \left. U_{i2}^*  V_{i1} {M_{\chi_i} \over m_b} {\lambda_b \over g_2}  
    \left\{P_{212}(x_i,x_i)+{2\over 3}
    P_{122}(x_i,x_i)\right\} (\d^u_{23})_{LL} \right] \ , \nn \\
C_9&=&(\d^u_{23})_{LR}  \fr{M^2_W}{M_{sq}^2 }\fr{2}{3}
   {\lambda_t \over g_2}{K_{cs}^*\over K_{ts}^* }\times\nn \\
   & &\sum_{i=1,2} V_{i2}^* V_{i1} \left\{ P_{312}(x_i,x_i)+
   \fr{1}{3} P_{042}(x_i,x_i)+ x_i P_{313}(x_i,x_i) \right\}, 
\nn \\
C'_7&=&(\d^u_{23})_{LR} \fr{M^2_W}{M_{sq}^2 }\fr{1}{3}
     {\lambda_t \over g_2}   {K_{cs}^* \over K_{ts}^*} \fr{m_s}{m_b}
    \sum_{i=1,2} V_{i2}^* V_{i1} \left\{\fr{3}{2} P_{222}(x_i,x_i)+
    P_{132}(x_i,x_i)\right\}\ . 
\eea

\subsection{Single mass insertion -- box}

Finally we compute the contributions which come from chargino box
diagrams of fig.~\ref{fig:box}. 

In the wino exchange case the result is  
\begin{equation}
C_9 = - C_{10} = (\d^u_{23})_{LL} {K_{cs}^*\over K_{ts}^*} {M_W^2 \over 
M_{sq}^2} {1\over \sw} \sum_{i,j=1,2} (V_{i1}^* V_{j1} V_{i1} V_{j1}^* )
f(x_i,x_j,x_{\tilde \nu}) \ .
\end{equation}
where
\begin{equation}
f(x_i,x_j,x_{\tilde \nu}) = {1\over 2}\int_0^1 dx \int_0^1 dy \int_0^1 dz 
{ yz(1-z)^2  \over \left[ y(1-z) + x_{\tilde \nu} (1-y)(1-z) +z (x_i
x+x_j (1-x)) \right]^2}
\end{equation}
and $x_{\tilde \nu} = M^2_{\tilde \nu} / M^2_{sq}$.

If the wino-bottom-stop vertex is replaced by an Higgsino-bottom-stop
one we obtain
\begin{equation}
C_9 = - C_{10} = -(\d^u_{23})_{LR} {K_{cs}^*\over K_{ts}^*} {M_W^2 \over 
M_{sq}^2} {\lambda_t \over g_2 \sw} \sum_{i,j=1,2} (V_{i1}^* V_{j1} 
V_{i1} V_{j2}^*) 
f(x_i,x_j,x_{\tilde \nu}). 
\end{equation}

\subsection{Double mass insertion -- Z}

It was recently pointed out~\cite{cola} that a double  mass insertion can
provide a great enhancement of the SUSY contribution to the decay
width, at least in the K-system case, if the $\d$s are not very constrained.

For B decay we obtain contributions from this graph to $C_9$ and $C_{10}$,
\bea
&&-\fr{C_9}{1-4\sw}=C_{10}=-{ (\d_{23}^u)_{LR} (\d_{33}^u)_{LR} \over 
4 \sw} 
{K_{cs}^* \over K_{ts}^*} \sum_{i,j=1,2} V_{i1} V_{j1}^*
\times \nn \\
&&
\left\{
 U^*_{i1} U_{j1}
\sqrt{x_{i} x_{j}} P_{123}(x_i,x_j) +\fr{1}{2} V^*_{i1} V_{j1}
 P_{122}(x_i,x_j)-
\fr{\d_{ij}}{3} P_{032}(x_i,x_j)
\right\} .\;
\label{eq:doubw}
\eea

\section{Gluino interactions}
\label{sec:gluins}
 The main contribution of  this kind of interactions comes from the graphs 
drawn in fig.~\ref{fig:sfigs}C,D.
In what follows we analyze the single and double mass insertion cases.

\subsection{Single mass insertion -- $\gamma$}
The corrections to the  coefficients in the photon mediated decay case
are:
\bea
C_7&=&\fr{\sqrt{2}}{M_{sq}^2 G_F} \ \fr{1}{3}\ \fr{N_c^2-1}{2N_c}
 {\pi \as \over K_{ts}^* K_{tb}}  
\left[\left( (\d^d_{23})_{LL}+(\d^d_{23})_{RR} \fr{m_s}{m_b}\right) 
       \fr{1}{4}P_{132}(x,x)+ (\d^{d}_{23})_{RL} P_{122}(x,x) \fr{M_{gl}}{m_b}\right],
 \nn \\
C_7^\prime&=& \fr{\sqrt{2}}{M_{sq}^2 G_F}\ \fr{1}{3}\ \fr{N_c^2-1}{2N_c}
  {\pi \as \over K_{ts}^* K_{tb}}  
\left[\left( (\d^d_{23})_{RR}+(\d^d_{23})_{LL} \fr{m_s}{m_b}\right) 
       \fr{1}{4}P_{132}(x,x)+ (\d^{d}_{23})_{LR} P_{122}(x,x) \fr{M_{gl}}{m_b}\right] \ ,
\nn \\
C_9 &=&
-\fr{\sqrt{2}}{M_{sq}^2 G_F}\ \fr{1}{3} \  \fr{N_c^2-1}{2N_c}
{\pi \as \over K_{ts}^* K_{tb}}   \fr{1}{3} 
P_{042}(x,x) (\d^d_{23})_{LL} \ ,
 \nn \\
C_9^\prime &=&
-\fr{\sqrt{2}}{M_{sq}^2 G_F}\ \fr{1}{3}\  \fr{N_c^2-1}{2N_c} 
{\pi \as \over K_{ts}^* K_{tb}}  \fr{1}{3} P_{042}(x,x) 
(\d^d_{23})_{RR} \ .
\label{eq:glu}
\eea

 The term  proportional to the gluino mass in eq.~(\ref{eq:glu}) seems to be strongly enhanced with
respect to the others. However the mass insertion which enters the
diagram is also strongly constrained from $b\rightarrow s \g$~\cite{gabb}.

\subsection{Single mass insertion -- Z}
The only relevant contributions to the Z-boson mediated decay width come from
diagrams in which the Z feels directly the breaking of $SU(2)_L$.  
According to the argument of  section~\ref{sec:opba} all the diagrams that 
do not respect this condition are suppressed with respect to the 
photon mediated ones and can  be neglected.
However, for penguins  containing a gluino,
 an explicit $SU(2)_L$ breaking can be provided 
 only with a double MI.
If only one MI is considered, Z-mediated  decays are completely negligible
with respect to the $\gamma$-mediated ones.

\subsection{Double  mass insertion -- Z}
For completeness we report here also the result obtained performing  a double mass
insertion in the gluino penguin.
\bea
&&-\fr{C_9}{1-4\sw}=C_{10} = \fr{(\d^d_{33})_{LR}(\d^d_{23})_{RL}}{K_{tb} K_{ts}^*} 
\fr{N_c^2-1}{2N_c} \fr{\as}{12  \a } \ P_{032}(x,x)\ , \nn \\
&&-\fr{C'_9}{1-4\sw}=C'_{10} = \fr{(\d^d_{33})_{RL}(\d^d_{23})_{LR}}{K_{tb} K_{ts}^*}
\fr{N_c^2-1}{2N_c} \fr{\as}{12  \a} \ P_{122}(x,x)\ .
\label{eq:gluz2}
\eea

\section{Light $\tilde t_R$ effects}
\label{sec:lightstop}
In the Mass Insertion Approximation framework we assume that all the
diagonal entries of the scalar mass matrices are degenerate and that
the off diagonal ones are sufficiently small. In this context we
expect all the squark masses to lie in a small region around an average
mass which we have chosen not smaller than 250 GeV. Actually there is
the possibility for the $\tilde t_R$ to be much lighter; in fact the
lower bound on its mass is about 70 GeV.
For this reason it is natural to wonder how good is the MIA when a
$\tilde t_R$ explicitly runs in a loop. 

The diagrams, among those we have computed, interested in this effect are 
the chargino penguins and box with the $(\delta^u_{23})_{LR}$
insertion. To compute the light--$\tilde t_R$ contribution we adopt
the approach presented in ref.~\cite{luca}. There the authors 
consider an expansion valid for unequal diagonal entries which
gives exactly the MIA in the limit of complete degeneration.  

The new expressions for the contributions to the coefficients
$C_{9}$ and $C_{10}$ are the following.
\begin{itemize}
\item Chargino Z--penguin with both an higgsino and a wino vertex:
\bea
&&- \fr{C_9}{1-4\sw}=C_{10}=  
(\d_{23}^u)_{LR} {\lambda_t \over g_2} {K_{cs}^*\over K_{ts}^* } 
{1\over 4 \sw}
\sum_{i,j=1,2} V_{i1} V_{j2}^* \times \nn \\
 &&
\left\{-U^*_{i1} U_{j1} \sqrt{x_{i} x_{j}} j(x_i,x_j,x_{\tilde t_R}) 
 + {1\over 2} V^*_{i1} V_{j1} k(x_i,x_j,x_{\tilde t_R})-
\fr{1}{2} \d_{ij} k(x_i,x_{\tilde t_R},1)
\right\} 
\label{eq:higwstop}
\eea
where $x_{\tilde t_R}=(m_{\tilde t_R} / M_{sq})^2$ and the functions
$j(x,y,z)$ and $k(x,y,z)$ can be found in ref~\cite{luca}.
\item Chargino $\gamma$--penguin with both an higgsino and a wino vertex:
\bea
C_9&=&(\d^u_{23})_{LR}  \fr{M^2_W}{M_{sq}^2 }\fr{2}{3}
   {\lambda_t \over g_2}{K_{cs}^* \over K_{ts}^* }
   \sum_{i=1,2}  V_{i1} V_{i2}^* \; g_7 (x_i,x_{\tilde t_R}) \ , \nn \\
C_7&=&(\d^u_{23})_{LR}  \fr{M^2_W}{M_{sq}^2 }\fr{1}{6}
   {\lambda_t \over g_2}{K_{cs}^* \over K_{ts}^* }
   \sum_{i=1,2} V_{i1} V^*_{i2} \; g_1 (x_i,x_{\tilde t_R}) 
\eea
where 
\bea
g_i (x,y) & = & {f_i(x) - {\ds 1\over \ds y} f_i({\ds x \over \ds y})\over 1-y} \ , \nn \\
f_7 (x)   & = &  {52 - 153 x + 144 x^2 - 43 x^3  + (36-54 x+12 x^3 )\log (x)\over 6(-1+x)^4}\ ,\nn \\
f_1 (x)   & = & {-8 + 3 x + 12 x^2 - 7 x^3  + 6 x (-3 +2x) \log (x) \over 6  (-1+x)^4}\ .
\eea
\item Chargino box with an higgsino vertex:
\beq
C_9 = - C_{10} = {1\over 4}(\d^u_{23})_{LR} {K_{cs}^*\over K_{ts}^*} {M_W^2 \over 
M_{sq}^2} {\lambda_t \over g_2 \sw} \sum_{i,j=1,2}  (V_{i1}^* V_{j1} 
V_{i1} V_{j2}^*) \;  k(x_i,x_j,x_{\tilde \nu},x_{\tilde t_R}) 
\eeq 
where $k(x,y,z,t)$ is defined in ref.~\cite{luca}.
\end{itemize}
All the above formulas reduce exactly to those presented
in sect.~\ref{sec:chains} in the limit $x_{\tilde t_R} = 1$.

\section{Constraints on mass insertions}
\label{sec:deltas}

In order to establish how large 
the SUSY contribution to $B\rightarrow X_s \ell^+ \ell^-$ can be, 
one can compare, coefficient
per coefficient, the MI results with the SM ones 
 taking into account possible constraints
on the $\d$s coming from other processes.

The most relevant $\d$s interested in the determination of the Wilson
coefficients $C_7$, $C_9$ and $C_{10}$ are
$(\d_{23}^u)_{LL}$, $(\d_{23}^u)_{LR}$, $(\d_{33}^u)_{RL}$,
$(\d_{23}^d)_{LL}$ and  $(\d_{23}^d)_{LR}$.

\begin{itemize}
\item
Vacuum stability arguments regarding the absence in the potential of
color and charge breaking minima and of directions unbounded from
below ~\cite{casa} give
\beq
(\d_{i3}^u)_{LR}\leq m_t {
\sqrt{2 M^2_{\tilde u}+2 M^2_{\tilde l}}
\over M_{sq}^2} \simeq 2 {m_t \over M_{sq}}.
\eeq
For  $M_{sq}\leq 300 $GeV this is not an effective constraint 
on the mass insertions.
\item
 A  constraint on $(\d_{23}^{d,u})_{LL}$
 can come from the possible measure of  $\Delta M_{B_s}$.

In fact the gluino--box contribution to
$\Delta M_{B_s}$~\cite{gab} is proportional to
$(\d_{23}^d)_{LL}^2$ (see for instance ref.~\cite{gab}).
A possible experimental determination of $\Delta M_{B_s}$, say
\beq
\Delta M_{B_s} < 30 \;  {\rm ps}^{-1}
\eeq
would imply that
\beq 
\sqrt{{\rm Re} (\d_{23}^d)_{LL}^2} < 0.5
\label{deltamslimit}
\end{equation}
for squark masses about $250 \gev$.
Moreover the LL up- and down-squark soft breaking mass matrices          
are related by  a  Cabibbo-Kobayashi-Maskawa rotation
\beq
(M^d_{sq})_{LL}^2= K^\dagger (M^u_{sq})_{LL}^2 K
\eeq
 so that the limit~(\ref{deltamslimit}) would be valid for the up sector too:
\beq
\sqrt{{\rm Re} (\d_{23}^u)_{LL}^2} < 0.5 \;.
\label{deltamslimitu}
\eeq
\item
Some  constraints come from the measure of  $B \rightarrow
X_s \g$. The branching ratio of this process depends almost completely
on the Wilson coefficients $C_7$ and $C_7^{\prime}$ which
are proportional respectively to 
$(\d_{23}^d)_{LR\  {\rm or}\ RL}$ and $(\d_{23}^u)_{LL}$.
The most recent CLEO estimate of the branching ratio for 
$B \rightarrow X_s \; \g$ is~\cite{cleo}
\begin{equation}
{\rm B}_{exp}(B\rightarrow s \ \g)=(3.15 \pm 0.35 \pm 0.32 \pm 0.26)\cdot 10^{-4}\ .
\end{equation}
where the first error is statistical, the second is systematic and the
third comes from the model dependence of the signal. The limits given at
95\% C.L. are~\cite{cleo}:
\beq
\begin{array}{ccccc}
 2.0 \ 10^{-4} &<& {\rm B}_{exp}(B\rightarrow s \ \g)  &< & 4.5 \ 10^{-4}. 
\end{array}
\label{limits}
\eeq

We can define a $C_7^{\rm eff}(M_B)$ as
\begin{equation}
\left| C_7^{\rm eff}(M_B)\right|^2={B_{exp}(B\rightarrow s \ \g) \over
\left(\ds K_{ts}^* K_{tb} \over \ds K_{cb}\right)^2  \fr{ \ds 6\a F}{ \ds \pi g(z)} }
\eeq
where $F$ can be found for instance in ref.~\cite{bsche}.
Considering the experimental limits given in eq.~(\ref{limits}) we find
\begin{equation}
0.28<|C_7^{\rm eff}(M_B)|<0.41.
\label{eq:cset}
\end{equation}

Actually $|C_7^{\rm eff}(M_B)|^2=|C_7 (M_B)|^2 + |C'_7(M_B)|^2$ and the
constraint given in eq.~(\ref{eq:cset}) should be shared between the two
coefficients. However in order to get the maximum SUSY contribution,
we observe that in physical observables $C'_7$ does not interfere with $C_7$, 
the $C'_7 C_9$ term is
suppressed by a factor  $m_s/m_b$ with respect to the $C_7 C_9$ one 
and $C'_7 C'_9$ is numerically negligible (in fact $C'_9$ is much
smaller than $C_9$). 
For these reasons we choose to fill the constraint of
eq.~(\ref{eq:cset}) with $C_7 (M_B)$ alone.

The bounds (\ref{eq:cset}) are referred to the coefficient evaluated at the
$M_B$ scale while we are interested to the limits at the much higher
matching scale.
After the RG evolution has been performed we find
that for an average
squark mass lower than 1 TeV, the MIA contribution alone  with a
suitable choice of $\delta$s, can always fit the experimental constraints.

Thus, since we are interested in computing the maximum enhancement
(suppression) SUSY can provide, we can 
choose the total $C_7^{eff}(M_B)$ anywhere inside the
allowed region given in eq.~(\ref{eq:cset}) still remaining 
consistent with the MIA.

The limit we get for $(\delta^d_{23})_{LR}$ is
of order $10^{-2}$ and this rules out Z-mediated
gluino penguins contributions to $C_9$ and $C_{10}$.

For what concerns  $(\delta^u_{23})_{LL}$ we find that the constraint
changes significantly according to the sign of $C_7^{eff}(M_B)$.
In this case it is important to consider both the positive and negative
region as this delta can
give a non negligible contribution  to $C_9$ and $C_{10}$.
The limits depend on the choice of the parameters in the chargino 
sector; the numerical results given below are computed for
$M_{sq} \simeq 250$ GeV, $\mu\simeq -160$ GeV, 
 $M_{\tilde \nu}\simeq 50$ GeV, $\tan \beta\simeq 2$
(in sect.~\ref{sec:results} we will show that 
these are the conditions under which we find the best SUSY 
contributions).
Considering the positive interval we find 
$-0.7<(\delta^u_{23})_{LL}<-0.5$ while in the negative one
$|(\delta^u_{23})_{LL}|<0.1$. 
\item
 Finally a comment on the  $\d$s  coming in  
 graphs with a  double MI is in order.

Given the  constraints on $(\delta^d_{23})_{LR}$  one can see that
 the gluino-penguins with a double MI give negligible contributions 
to the final results even if  $(\delta^d_{33})_{RL}$
 is of order ${\cal O}(1)$.

A $(\delta^u_{33})_{RL}$  of order ${\cal O}(1)$,   can give rise to
light  or negative squark mass eigenstates.
In particular a light $\tilde{t}_L$ would
contribute too much to the $\rho_W$-parameter.
Eventual model dependent cancelation can provide an escape to these
constraints.
In any case  the numerical value of these contributions
is not particularly  important for the determination of physical observables.
Since we want to provide a model independent analysis we prefer 
not to consider in our final computation these double insertion graphs  and we
present them only for  completeness.

Contributions with three mass insertions are suppressed due to           
small loop integrals and to the various constraints on the deltas. 
\end{itemize}

\section{Results}
\label{sec:results}

\begin{figure}
\begin{center}

\vspace*{-4.5cm}
\epsfig{file=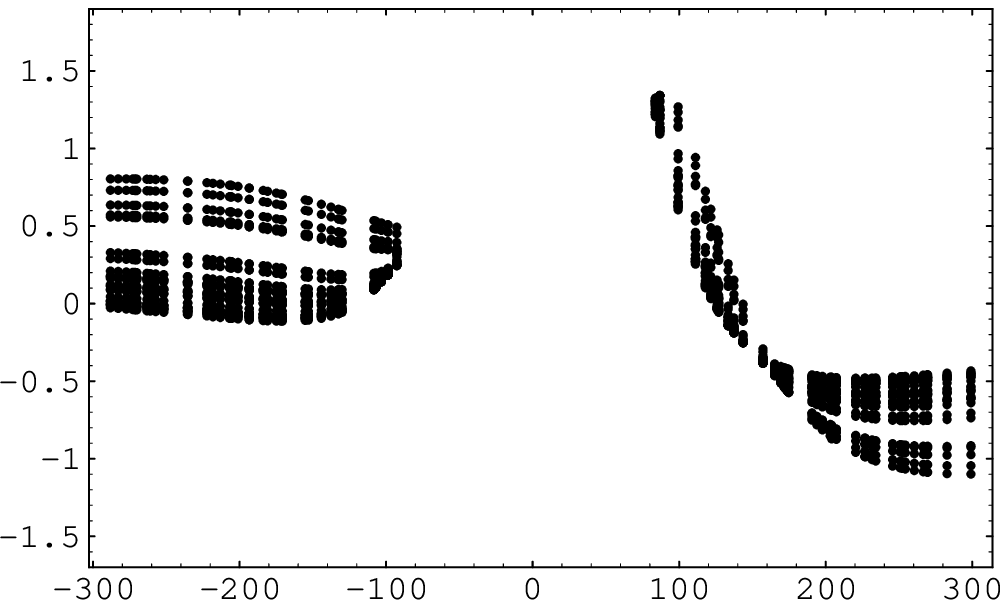,width=0.9\linewidth}

\vspace*{-5cm}
\epsfig{file=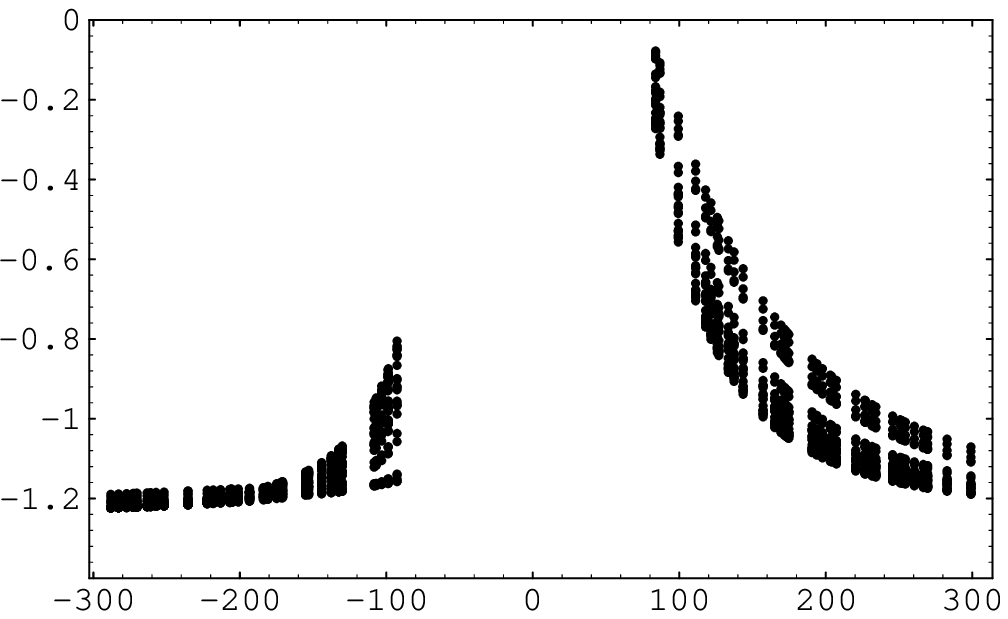,width=0.9\linewidth}

\vspace*{-2cm}
\caption[]{\it $(\d_{23}^u)_{LR}$ (above) and $(\d_{23}^u)_{LL}$
(below) contributions to $C_{9}$ coming from
chargino diagrams as functions of $\mu$ (expressed in GeV). $M_{sq}$
is fixed to 250 GeV while $\tan \beta$ varies between 2 and 30.}
\protect\label{fig:scatt9}
\end{center}
\end{figure}

\begin{figure}
\begin{center}

\vspace*{-4.5cm}
\epsfig{file=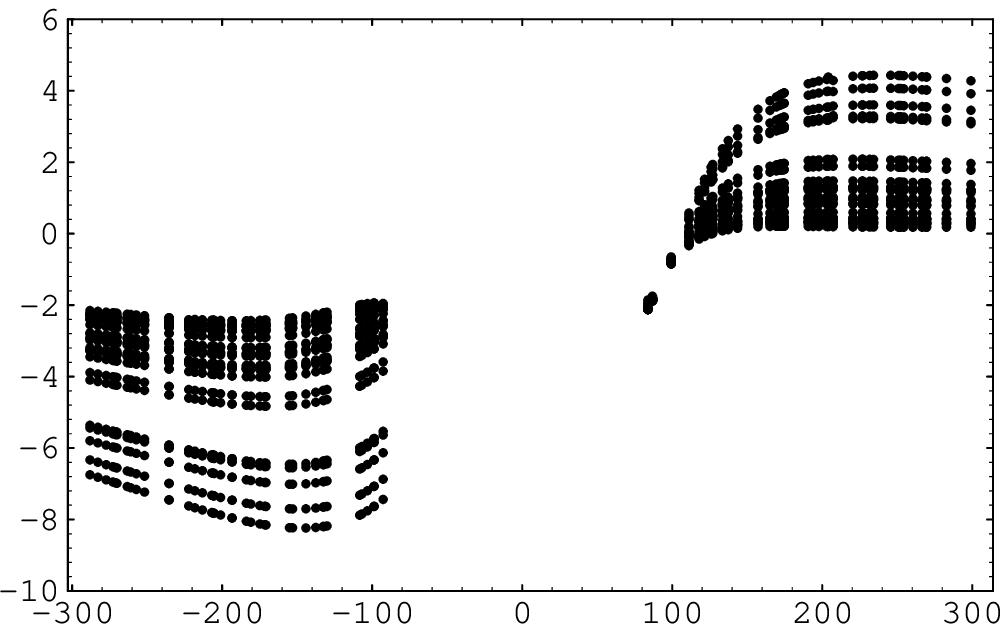,width=0.9\linewidth}

\vspace*{-5cm}
\epsfig{file=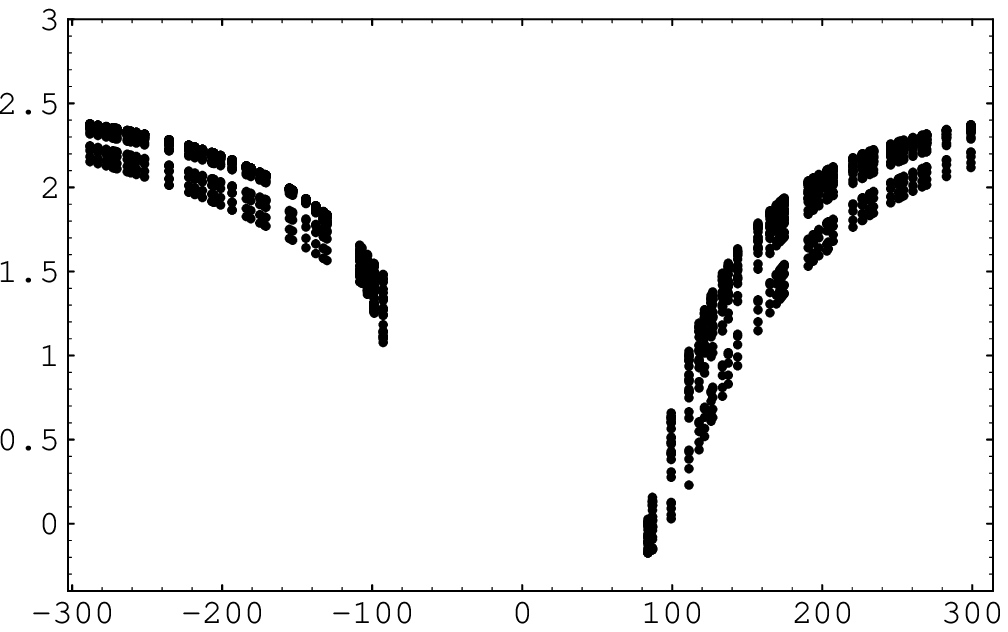,width=0.9\linewidth}

\vspace*{-2cm}
\caption[]{\it $(\d_{23}^u)_{LR}$ (above) and $(\d_{23}^u)_{LL}$
(below) contributions to $C_{10}$ coming from
chargino diagrams as functions of $\mu$ (expressed in GeV). $M_{sq}$
is fixed to 250 GeV while $\tan \beta$ varies between 2 and 30.}
\protect\label{fig:scatt10}
\end{center}
\end{figure}

The results of the calculations of sections~\ref{sec:chains}-\ref{sec:gluins}
are presented in figs.~\ref{fig:scatt9}-\ref{fig:scatt10} and in
tables~\ref{tab:gl}-\ref{tab:ch}. 
While the gluino sector of the theory is essentially determined by the
knowledge of the gluino mass (i.e.~$M_{gl}$), the chargino one needs
two more parameters (i.e.~$M_2$, $\mu$ and $\tan \beta$). 
Moreover it is
 a general feature of the models we are studying the decoupling of
the SUSY contributions in the limit of high sparticle masses:
we expect the biggest SUSY contributions to appear for such masses
chosen at the lower bound of the experimentally allowed region.
On the other hand this considerations  suggest us to constrain the
three parameters of the chargino sector by the requirement of the lighter eigenstate not
to have a mass lower than  the experimental bound of  about 70 GeV~\cite{pdg}. 
 The  remaining two dimensional space has  yet no  constraint. 
For these reasons we scan the chargino parameter space by means of
scatter plots for which $M_{sq} = 250 \; \hbox{GeV}$, $M_{\tilde \nu} =
50 \; \hbox{GeV}$, $80 \; \hbox{GeV} \leq |\mu | \leq 300 \;
\hbox{GeV}$ and $2 \leq \tan \beta \leq 30$; 
for every choice of these two parameters, $M_2$ is
determined imposing to the lighter eigenstate a mass of about 70 GeV.
In the plots we sum  all contributions coming from different graphs 
  proportional 
 to a common  mass insertion (the actual values of the coefficients
are obtained multiplying the points in the plots by the MI).

In the tables we report the contribution of each diagram and 
the explicit dependence on the mass insertion parameters.
We evaluate the coefficients varying $ \ M_{sq}$ and $ M_{gl}$
between 250 GeV and 1 TeV.
The other parameters  in tab.~\ref{tab:ch} are fixed from the
scatter plots in order to give the best SUSY contributions to $C_9$ 
and $C_{10}$. 

Thus, with $\mu\simeq -160$, $M_{gl}\simeq M_{sq} \simeq 250$ GeV,
 $M_{\tilde \nu}\simeq 50$ GeV, $M_{\tilde t_R}=90 \gev$, $\tan \beta\simeq 2$
 one gets
\beq
\cases{C_9^{MI}(M_B) = -1.2 \deu_{LL} + 0.69 \deu_{LR} -0.51(\delta^d_{23})_{LL}      &\cr    
      C_{10}^{MI}(M_B) =  1.75 \deu_{LL} - 8.25 \deu_{LR} \ .   &\cr}
\label{cout}
\end{equation}

In order to numerically compare  eq.~(\ref{cout}) with  the respective SM 
results we note that the minimum value of
 $\left(C_9^{\rm eff}(s)\right)^{SM}(M_B)$ is about 4 
while $C_{10}^{SM} = -4.6$. 
Thus one   deduces that 
SM expectations  for the observables  are 
 enhanced when $C_9^{MI}(M_b)$ is positive.
 Moreover  the big value of   $C_{10}^{MI}(M_B)$  implies that
the final  total coefficient $C_{10}(M_B)$ can  have a different sign with 
respect to the SM estimate.
As a  consequence of this,
 the sign of   asymmetries can  be the opposite of the
one calculated in the SM.

The diagonal contributions to $C_9$, $C_{10}$ introduced in 
sect.\ref{sec:opba}, and  computed in the same range of the parameters 
(with $M_{H^+}$ chosen just above the experimental threshold of about 
$60 \gev$) are
\beq
\cases{C_9^{diag}(M_B) = -0.35 &\cr
       C_{10}^{diag}(M_B)= -0.27 &\cr}
\end{equation}

The sign and the value of  the coefficient $C_7$ has a great
importance.
In fact the  integral  of the BR (see eq.~(\ref{eq:br0})) is dominated
 by the $|C_7|^2/s $  and  $C_7  C_9$ term  for low values of $s$.
In the SM the interference between $O_7$ and $O_9$ is destructive and
this behavior can be easily modifed in the general class of models we
are dealing with.

In the following, according
to  the discussion of sect.~\ref{sec:deltas},
 we give the configurations of the various $\delta$s for which we
find the best enhancements and suppressions of the SM expectations.
\begin{itemize}
\item {\bf Best enhancement.} 

\begin{tabular}{|c|c|c|c||c|c|} \hline
             &$C_7^{eff}(M_B)$&$C^{MI}_9$&$C^{MI}_{10}$&$(\delta^{u,d}_{23})_{LL}$&$(\delta^u_{23})_{LR}$\\ \hline  
$R$                & 0.41 & 1.5  & -8.3 & -0.5 & 0.9  \\ \hline 
$A_{FB}$	   & 0.41 & 0.96 & -2.1 & -0.5 & 0.15 \\ \hline
$\overline{A}_{FB}$& 0.28 & 0.96 & -2.1 & -0.5 & 0.15 \\ \hline 
\end{tabular} 

It is important to note that with such choices
the behavior of the asymmetries in the low $s$ region
of the spectrum is greatly modified: the coefficients of the operators
$Q_7$ and  $Q_9$ sum up instead of cancel each other in such a way
that the asymmetries are never negative. It is also important to
stress that the asymmetries get their extremal value with a rather
small $(\delta^u_{23})_{LR}$: the enhancement given here will survive possible
future constraints on this insertion.
\item {\bf Best enhancement with $\mathbf C_7 <0$.} 

\begin{tabular}{|c|c|c|c||c|c|} \hline
             &$C_7^{eff}(M_B)$&$C^{MI}_9$&$C^{MI}_{10}$&$(\delta^{u,d}_{23})_{LL}$&$(\delta^u_{23})_{LR}$\\ \hline  
$R$,      &-0.41&1.5   &-8.3    &-0.5            &0.9	\\ \hline 
$A_{FB}$, $\overline{A}_{FB}$ &-0.28&0.75  &0.36    &-0.5 &-0.15	\\ \hline 
\end{tabular}
\item {\bf Best depression.} 

\begin{tabular}{|c|c|c|c||c|c|} \hline
             &$C_7^{eff}(M_B)$&$C^{MI}_9$&$C^{MI}_{10}$&$(\delta^{u,d}_{23})_{LL}$&$(\delta^u_{23})_{LR}$\\ \hline  
$R$      &-0.28&-1.3  &5.8  &0.5            &-0.6	\\ \hline 
 $A_{FB}$, $\overline{A}_{FB}$&0.28&-1.5  &8.3    &0.5    &-0.9	\\ \hline 
\end{tabular} 
\end{itemize}
\begin{table}
\begin{center}
\begin{tabular}{||c||c||c|c||c|c||c|c||} \hline
Observable & SM & SUSY    & SUSY/        &SUSY    & SUSY/     & SUSY & SUSY/   \\  
           &    & maximal &    SM    &minimal &      SM   & ($C_7<0$) &SM\\ \hline \hline
\vphantom{\fbox{$BR (e)$}}$BR (e)$&$9.6\,{{10}^{-6}}$&$4.4 \
10^{-5}$&$4.6$&$3.9\,{{10}^{-6}}$&$0.41$&$3.9 \ 10^{-5}$&$4.0$ \\  \hl 
\vphantom{\fbox{$BR (e)$}}$A_{FB} (e)$&$0.23$&$0.33$&$1.5$&$-0.18$&$-0.78$&$0.31$&$1.4$ \\ \hl 
\vphantom{\fbox{$BR (e)$}}$\overline A_{FB} (e)$&$0.071$&$0.24$&$3.3$&$-0.19$&$-2.7$&$0.11$&$1.5$  \\ \hline \hline 
\vphantom{\fbox{$BR (e)$}}$BR (\mu)$&$6.3\,{{10}^{-6}}$&$4.0 \ 10^{-5}
$&$6.3$&$1.6\,{{10}^{-6}}$&$0.26$&$3.4\ 10^{-5} $&$5.4$  \\ \hl 
\vphantom{\fbox{$BR (e)$}}$A_{FB} (\mu)$&$0.23$&$0.33$&$1.5$&$-0.18$&$-0.78$&$0.31$&$1.4$  \\ \hl 
\vphantom{\fbox{$BR (e)$}}$\overline A_{FB} (\mu)$&$0.11$&$0.27$&$2.5$&$-0.27$&$-2.4$&$0.15$&$1.3$ \\ \hl 

\end{tabular}
\caption{\it Integrated BR, $A_{FB}$ and $\overline A_{FB}$ in
the SM and in a general SUSY extension of the SM for the decays
$B \rightarrow X_s e^+ e^-$ and $B \rightarrow X_s \mu^+ \mu^-$. 
The second and third columns
are the extremal values we obtain with a positive $C_7^{eff}$ while
the fourth one is the $C_7^{eff}<0$ case. 
The actual numerical inputs
for the various coefficients can be found in the text.
The BR is just the integral of $R(s)$ multiplied by the BR of the
semileptonic dominant B decay ($BR(B \rightarrow X_c e \nu) = 0.105$). }
\label{tab:resinteg}
\end{center}
\end{table}

\begin{figure}
\vskip-6cm
\begin{center}
\epsfig{file=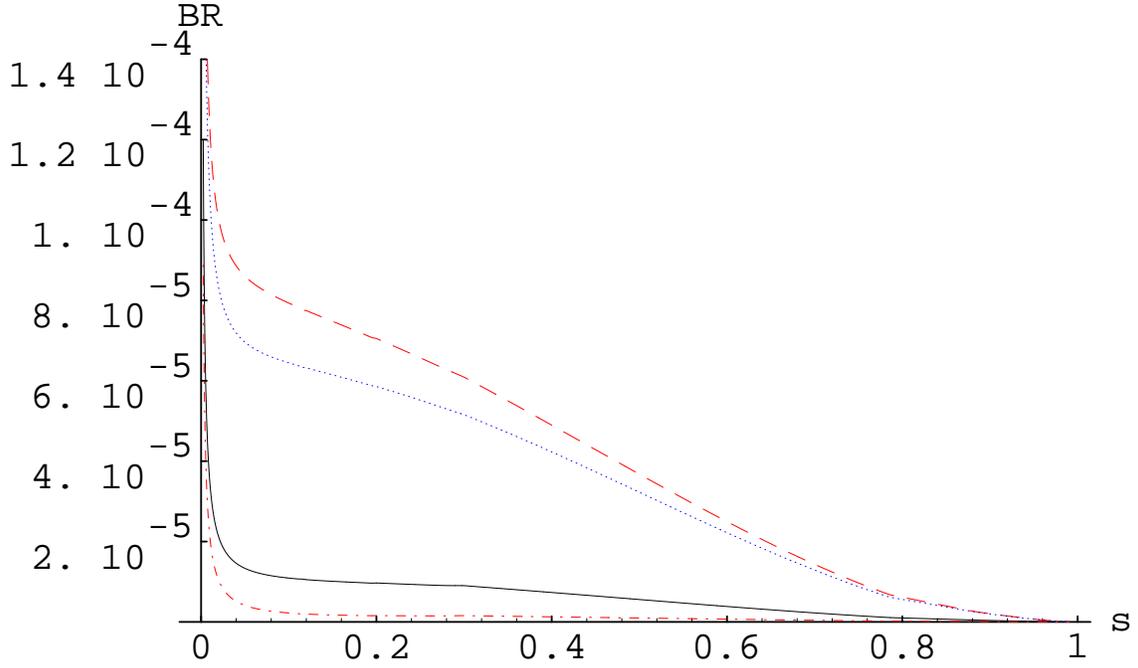,width=0.9\linewidth}
\vskip-2cm
\caption[]{\it Differential branching ratio  for the decay 
$B \rightarrow X_s \ell^+ \ell^-$. 
The solid line corresponds to the SM expectation; the dashed and
dotted--dashed lines correspond respectively to the SUSY best
enhancement ($C_7^{eff}=0.41$,$C_9^{MI}=1.5,C_{10}^{MI}=-8.3$) and
depression ($C_7^{eff}=-0.28$,$C_9^{MI}=-1.3,C_{10}^{MI}=5.8$);
the dotted line is the maximum enhancement obtained without 
changing the sign of $C_7$ ($C_7^{eff}=-0.41$,$C_9^{MI}=1.5,C_{10}^{MI}=-8.3$).}
\protect\label{fig:br}
\end{center}
\end{figure}

\begin{figure}
\begin{center}
\epsfig{file=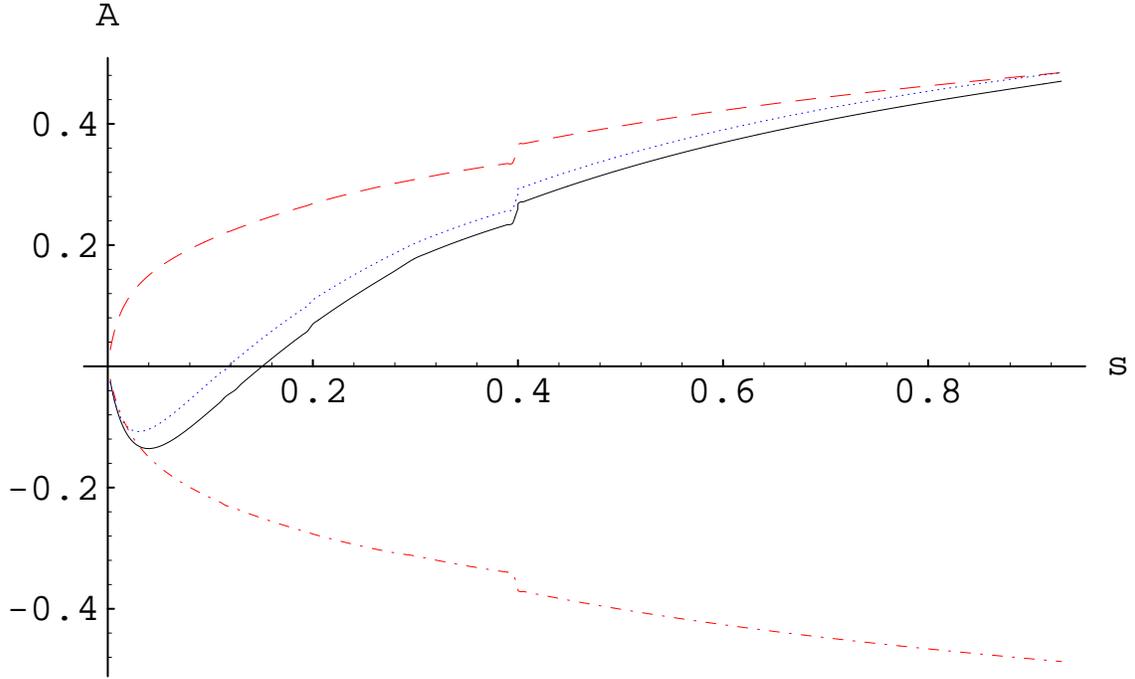,width=0.9\linewidth}
\caption[]{\it Forward--Backward asymmetry ($A_{FB}$) for the decay 
$B \rightarrow X_s \ell^+ \ell^-$.
The solid line corresponds to the SM expectation; the dashed and dotted-dashed
line  corresponds to the SUSY best
enhancement ($C_7^{eff}=0.41$,$C_9^{MI}=0.96,C_{10}^{MI}=-2.1$) and
depression ($C_7^{eff}=.28$,$C_9^{MI}=-1.5,C_{10}^{MI}=8.3$);
the dotted line is the maximum enhancement obtained without 
changing the sign of $C_7$ ($C_7^{eff}=-0.28$,$C_9^{MI}=0.75,C_{10}^{MI}=0.36$).}
\protect\label{fig:afb}
\end{center}
\end{figure}

\begin{figure}
\begin{center}
%\hspace*{-5cm}
\epsfig{file=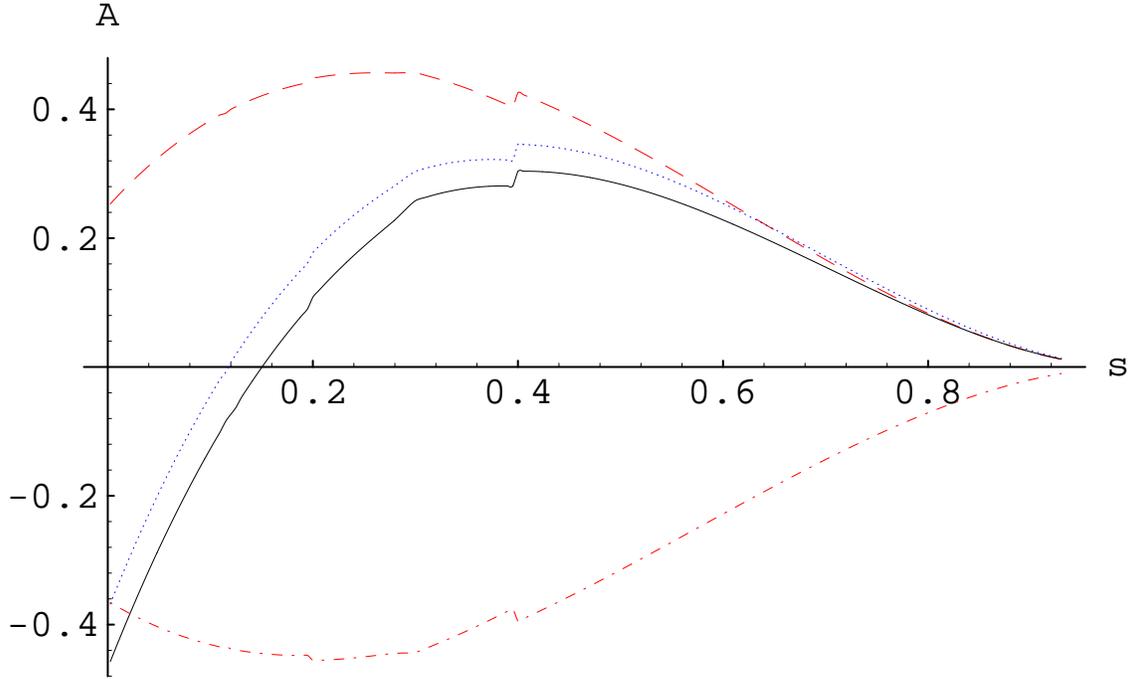,width=0.9\linewidth}
\caption[]{ \it Forward--Backward asymmetry ($\overline A_{FB}$) for the decay 
$B \rightarrow X_s \mu^+ \mu^-$. 
The solid line corresponds to the SM expectation; the dashed and
dotted--dashed lines correspond respectively to the SUSY best
enhancement ($C_7^{eff}=0.28$,$C_9^{MI}=0.96,C_{10}^{MI}=-2.1$) and
depression ($C_7^{eff}=0.28$,$C_9^{MI}=-1.5,C_{10}^{MI}=8.3$);
the dotted line is the maximum enhancement obtained without 
changing the sign of $C_7$ ($C_7^{eff}=-0.28$,$C_9^{MI}=0.75,C_{10}^{MI}=0.36$).}
\protect\label{fig:achomu}
\end{center}
\end{figure}

\begin{figure}
\begin{center}
%\hspace*{-5cm}
\epsfig{file=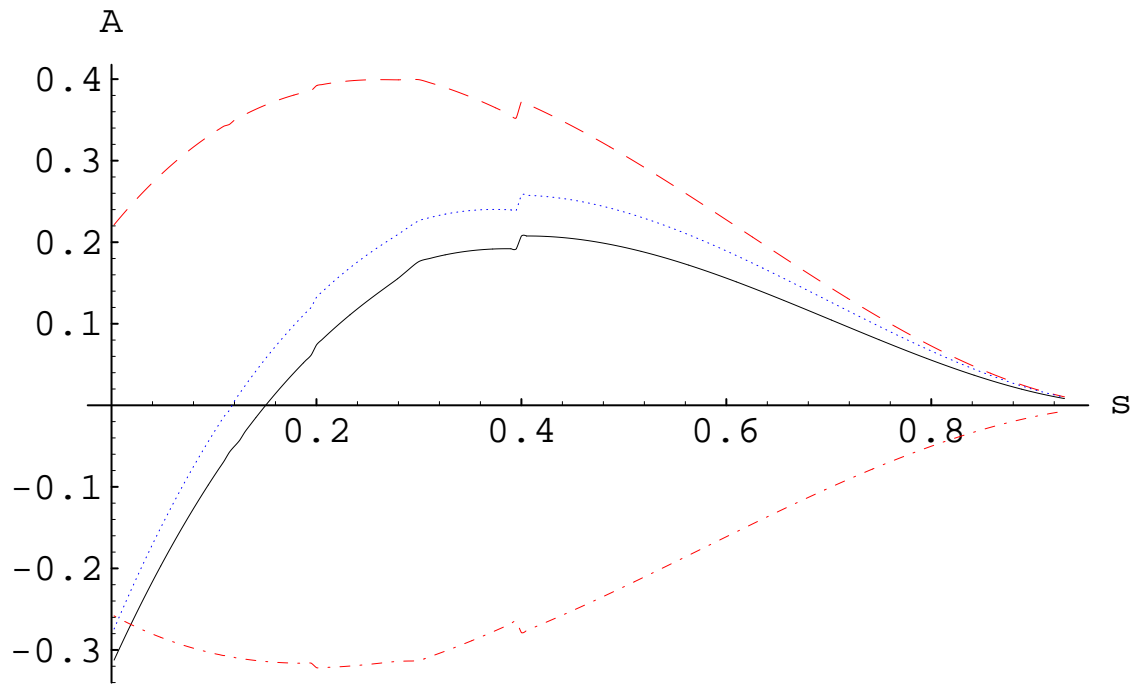,width=0.9\linewidth}
\caption[]{\it  Forward--Backward asymmetry ($\overline A_{FB}$) for the decay 
$B_d \rightarrow X_s e^+ e^-$ (See the caption in
fig.~\ref{fig:achomu}).} 
\protect\label{fig:achoe}
\end{center}
\end{figure}

The plots of $BR(s)$, $A_{FB}(s)$ and $\overline A_{FB}(s)$ 
are drawn in figs.\ref{fig:br}-\ref{fig:achoe}. 
Here both SM  and  SUSY results are shown. 
The discontinuity in the $A_{FB}$ plot at $s=0.7$  
corresponds  to the point at which we have stopped the corrections
$O(1/m_b^2)$. In fact a model independent description of the
differential asymmetry in the region 
$0.7<s<0.93$ beyond the parton model is still lacking.
Further the peak wich occur at $s=(2 m_c / m_b)^2 \simeq 0.3$
is due to the perturbative remnant of the $c \bar c$    
resonance.

The integrated BRs and asymmetries for the decays 
$B\rightarrow X_s e^+ e^-$ and 
$B\rightarrow X_s \mu^+ \mu^-$ in the SM case
and in the SUSY one (with the above choices of the parameters) are
summarized in tab.\ref{tab:resinteg}.
There we computed the total perturbative contributions neglecting the
resonances; these occur in the
intermediate range of the spectrum ($J/\psi$ at 3.1 GeV 
($s=0.42$) and  $\psi'$ at
3.7 GeV ($s=0.59$) plus others at higher energies). However it is possible 
to exclude the resonant regions from the experimental analysis by 
opportune cuts and to correct the effects of their tails in the remaining
part of the spectrum. 

The results of tab.~\ref{tab:resinteg} must be compared with 
the experimental best limit which reads~\cite{cleodue}
\begin{eqnarray}
BR_{exp}&<& 5.8 \; 10^{-5}.
\end{eqnarray}
 
A comment  on the CMSSM
(Constrained MSSM) prediction for the observables we have computed is now
 necessary. 
An analysis on the subject is presented in ref.~\cite{cho}.
In this paper the authors show that the effect 
of CMSSM on the integrated BRs, 
considering only contributions to $C_9$ and $C_{10}$, 
varies between a depression up to 10\% and an
enhancement of few percents  relative to the corresponding SM values.
The asymmetries get even smaller corrections. 
On the other hand a direct computation of
$C_7^{MSSM}(M_W)$ yields~\cite{cho}
\begin{eqnarray}
-0.59 < C_7^{MSSM}(M_W) < +0.49 & \;\; & \hbox{in the large $\tan{\beta}$ regime}, \nonumber \\
-0.26 < C_7^{MSSM}(M_W) < -0.20 & \;\; & \hbox{in the low $\tan{\beta}$ regime}.
\end{eqnarray}
It is worth noting that comparing the above intervals with the
experimentally allowed region obtained via RG evolution at the $M_W$
scale of the limits in eqn.~(\ref{eq:cset}) (we use only the SM contribution to
$C_8$; the inclusion of the MSSM corrections does not change
significantly the result) 
\begin{equation}
-0.39 < C_7 (M_W) < -0.099 \; {\rm and} \;  0.66 < C_7 (M_W) < 0.95
\label{cset}
\end{equation}
it is excluded
that the CMSSM could drive a positive value for $C_7^{eff} (M_B)$.
For what concerns the negative interval of values of  $C_7^{eff}
(M_B)$ we see that 
it can be accommodated both in the CMSSM and in our framework.

Looking at figs~\ref{fig:br}-\ref{fig:achoe} and table~\ref{tab:resinteg}
 we see                 
that the differences between SM and SUSY predictions can be remarkable.
 Moreover a sufficiently precise measure of  BRs, $A_{FB}$s and
$\overline A_{FB}$s
can either discriminate between the  CMSSM and
more general SUSY models or give new constraints on mass insertions.
Both these kind of informations can be very useful for model building. 
%----------------------------------------------------------------------------
\begin{table}
\begin{center}
\begin{tabular}{|c|c|c|c|c|}  \hl
Diagram & $M_{sq}$ & $M_{gl}$ & $C_7$     & $C_9$   \spazio  \\ \hl
& 250 & 250 &$-0.192 (\delta^d_{23})_{LL} - 33.4 (\delta^d_{23})_{LR}$&$-0.513 (\delta^d_{23})_{LL}$ \spazio \\ 
& 250 & 500 &$-0.125 (\delta^d_{23})_{LL} - 31.2 (\delta^d_{23})_{LR}$&$-0.189 (\delta^d_{23})_{LL}$  \spazio \\ 
$\tilde{g} \gamma -{\rm 1 ins}$&500&500&$-0.0449 (\delta^d_{23})_{LL} - 15.6 (\delta^d_{23})_{LR}$&$-0.12 (\delta^d_{23})_{LL}$  \spazio \\ 
& 250  & 1000 &$-0.0344 (\delta^d_{23})_{LL} - 10.3 (\delta^d_{23})_{LR}$&$-0.0463 (\delta^d_{23})_{LL}$  \spazio \\ 
& 500  & 1000 &$-0.0291 (\delta^d_{23})_{LL} - 14.5 (\delta^d_{23})_{LR}$&$-0.0439 (\delta^d_{23})_{LL}$  \spazio \\ 
& 1000 & 1000 &$-0.0105 (\delta^d_{23})_{LL} - 7.26 (\delta^d_{23})_{LR}$&$-0.0279 (\delta^d_{23})_{LL}$ \spazio \\ \hline\hline  
       &    &      & $C_{10}$  & $C_9$   \spazio  \\ \hl  
& 250 & 250 &$-10.2 (\delta^d_{23})_{LR} (\delta^d_{33})_{RL}$&$0.763 (\delta^d_{23})_{LR} (\delta^d_{33})_{RL}$ \spazio \\ 
& 250 & 500 &$-17.3 (\delta^d_{23})_{LR} (\delta^d_{33})_{RL}$&$1.29 (\delta^d_{23})_{LR} (\delta^d_{33})_{RL}$  \spazio \\ 
$\tilde{g} Z -{\rm 2 ins}$&500&500&$-9.49 (\delta^d_{23})_{LR} (\delta^d_{33})_{RL}$&$0.712 (\delta^d_{23})_{LR} (\delta^d_{33})_{RL}$  \spazio \\ 
& 250  & 1000 &$-17.6 (\delta^d_{23})_{LR} (\delta^d_{33})_{RL}$&$1.32 (\delta^d_{23})_{LR} (\delta^d_{33})_{RL}$  \spazio \\ 
& 500  & 1000 &$-16.1 (\delta^d_{23})_{LR} (\delta^d_{33})_{RL}$&$1.21 (\delta^d_{23})_{LR} (\delta^d_{33})_{RL}$  \spazio \\ 
& 1000 & 1000 &$-8.85 (\delta^d_{23})_{LR} (\delta^d_{33})_{RL}$&$0.664 (\delta^d_{23})_{LR} (\delta^d_{33})_{RL}$  \spazio \\ \hl  

\end{tabular}
\caption{Contributions to the coefficients $C_7$, $C_9$ and $C_{10}$
from diagrams involving gluino loops. $M_{gl}$ and $M_{sq}$ both vary 
between 250 GeV and 1000 GeV. Exchanging L with R in the mass
insertions we get the contributions of gluino diagrams to $C'_7$,
$C'_9$ and $C'_{10}$. For further explanations see the caption in 
tab.~\ref{tab:ch}.}
\label{tab:gl}
\end{center}
\end{table}
%----------------------------------------------------------------------------
\begin{table}
\begin{center}
\begin{tabular}{|c|c|c|c|}  \hl
Diagram & $M_{sq}$ & $C_7$     & $C_9$   \spazio  \\ \hl
 & 250  &$0.35 (\delta^u_{23})_{LL}$&$1.4 (\delta^u_{23})_{LL}$  \spazio \\ 
$ \tilde{W} \tilde{W} \gamma -{\rm 1 ins}$ & 500  &$0.12 (\delta^u_{23})_{LL}$&$0.76 (\delta^u_{23})_{LL}$  \spazio \\ 
 & 1000 &$0.033 (\delta^u_{23})_{LL}$&$0.32 (\delta^u_{23})_{LL}$ \spazio \\ \hl 
 & 250  &$-2.1 (\delta^u_{23})_{LL} - 0.25 (\delta^u_{23})_{LR}$&$-0.71 (\delta^u_{23})_{LR}$  \spazio \\ 
$ \tilde{H} \tilde{W} \gamma -{\rm 1 ins}$ & 500  &$-1.1 (\delta^u_{23})_{LL} - 0.27 (\delta^u_{23})_{LR}$&$-0.87 (\delta^u_{23})_{LR}$  \spazio \\ 
 & 1000 &$-0.45 (\delta^u_{23})_{LL} - 0.27 (\delta^u_{23})_{LR}$&$-0.93 (\delta^u_{23})_{LR}$\spazio \\ \hline \hline 
       &          & $C_{10}$  & $C_9$   \spazio  \\ \hl  
 & 250  &$1.4 (\delta^u_{23})_{LR} (\delta^u_{33})_{RL}$&$-0.092 (\delta^u_{23})_{LR} (\delta^u_{33})_{RL}$  \spazio \\ 
$ \tilde{W} \tilde{W} Z -{\rm 2 ins}$ & 500  &$1.8 (\delta^u_{23})_{LR} (\delta^u_{33})_{RL}$&$-0.12 (\delta^u_{23})_{LR} (\delta^u_{33})_{RL}$  \spazio \\ 
 & 1000 &$2.1 (\delta^u_{23})_{LR} (\delta^u_{33})_{RL}$&$-0.14 (\delta^u_{23})_{LR} (\delta^u_{33})_{RL}$ \spazio\\ \hl
 & 250  &$-8.4 (\delta^u_{23})_{LR}$&$0.56 (\delta^u_{23})_{LR}$  \spazio \\ 
$ \tilde{W} \tilde{H} Z -{\rm 1 ins}$ & 500  &$-11. (\delta^u_{23})_{LR}$&$0.74 (\delta^u_{23})_{LR}$  \spazio \\ 
 & 1000 &$-13. (\delta^u_{23})_{LR}$&$0.84 (\delta^u_{23})_{LR}$ \spazio\\ \hl
 & 250  &$-0.91 (\delta^u_{23})_{LL}$&$0.06 (\delta^u_{23})_{LL}$  \spazio \\ 
$ \tilde{W} \tilde{W} Z -{\rm 1 ins}$ & 500  &$-0.47 (\delta^u_{23})_{LL}$&$0.031 (\delta^u_{23})_{LL}$  \spazio \\ 
 & 1000 &$-0.19 (\delta^u_{23})_{LL}$&$0.013 (\delta^u_{23})_{LL}$ \spazio\\ \hl
 & 250  &$2.7 (\delta^u_{23})_{LL}$&$-2.7 (\delta^u_{23})_{LL}$  \spazio \\ 
$\hbox{box} \tilde{W}  -{\rm 1 ins}$ & 500  &$1.3 (\delta^u_{23})_{LL}$&$-1.3 (\delta^u_{23})_{LL}$  \spazio \\ 
 & 1000 &$0.55 (\delta^u_{23})_{LL}$&$-0.55 (\delta^u_{23})_{LL}$ \spazio\\ \hl
 & 250  &$-0.97 (\delta^u_{23})_{LR}$&$0.97 (\delta^u_{23})_{LR}$  \spazio \\ 
$\hbox{box} \tilde{H} \tilde{W}  -{\rm 1 ins}$ & 500  &$-1.1 (\delta^u_{23})_{LR}$&$1.1 (\delta^u_{23})_{LR}$  \spazio \\ 
 & 1000 &$-1.2 (\delta^u_{23})_{LR}$&$1.2 (\delta^u_{23})_{LR}$ \spazio\\ \hl

\end{tabular}
\caption{Contributions to the coefficients $C_7$, $C_9$ and $C_{10}$ 
from diagrams involving chargino loops. We assume $\mu = -160$ GeV, 
$M_2 = 50$  GeV, $\tan \beta = 2$, $m_{\tilde \nu}=50$ GeV,
$M_{\tilde t_R}=90$ GeV  while $M_{sq}$ varies between 250 GeV and
1000 GeV. In the first column we indicate the number of mass insertions
present in each squark line, which charginos are present at the
vertexes and the kind of graph computed ($\gamma$--penguin,
$Z$--penguin or box diagram).}
\label{tab:ch}
 \end{center}
\end{table}

%----------------------------------------------------------------------------
%___________________________________________________________________________

%___________________________________________________________________________
%___________________________________________________________________________
%___________________________________________________________________________
%___________________________________________________________________________
%___________________________________________________________________________

\newpage
\section{Conclusions}
\label{sec:conclusions}
 In this paper an  extensive discussion about  SUSY contributions to
 semileptonic decays $B\rightarrow X_s e^+ e^-$, 
$B\rightarrow X_s \mu^+ \mu^-$ 
 is provided.
We see that the interplay between 
$b\rightarrow s \g$ and $B\rightarrow X_s \ell^+ \ell^-$ 
 is fundamental in order to give  an  estimate
 of the SUSY relevance in  these decays.
The two kinds of decays are in fact strongly correlated.

Given the constraints coming from  the 
 recent  measure of $b\rightarrow s \g$
 and estimating all possible SUSY effects in the MIA framework 
 we see that SUSY   has  a  chance  to strongly
enhance or depress semileptonic charmless B-decays.
The  expected direct measure will give very interesting informations
about the SM and its possible extensions.

\section*{Acknowledgments}
We thank S. Bertolini and E. Nardi for fruitful discussions.
I.S. wants to thank SISSA, for support and kind hospitality 
 during the elaboration of the first part of this work and Della Riccia Foundation
(Florence, Italy) for partial support.
The work of L.S. was supported by the German Bundesministerium f\"ur
Bildung und Forschung under contract 06 TM 874 and by the DFG project Li
519/2-2. 
This work was partially supported by INFN and by the TMR--EEC 
network ``Beyond the Standard Model'' (contract number ERBFMRX
CT960090).

\end{document}